\documentclass[pdr,
notitlepage,showpacs,preprintnumbers,amsmath,amssymb,nofootinbib,APS,8pt,superscriptaddress]{revtex4-1}

\usepackage{dcolumn}
\usepackage{bm}
\usepackage{xcolor}
\usepackage[utf8]{inputenc}
\usepackage[spanish,english]{babel}
\usepackage{amsmath,amssymb,amsfonts,latexsym,cancel}
\usepackage{graphicx}
\usepackage{color}
\usepackage{soul}
\usepackage{ulem}
\usepackage{hyperref}
\usepackage{amsmath}
\usepackage{slashed}
\usepackage{float} 
\usepackage{braket}

\begin{document}

%
%

\title{{\bf The hyperbolic Einstein-Rosen bridge} }

\author{Pau Beltr\'an-Palau}\email{pau.beltran@uv.es}
\affiliation{Departament de Física Teòrica and IFIC, Centro Mixto Universitat de València-CSIC. Facultat de Física, Universitat de València, Burjassot-46100, València, Spain.}
\author{Miguel Portilla}\email{miguel.portilla@uv.es}
\affiliation{Departament d'Astronomia i Astrofísica, Universitat de València, Burjassot-46100, València, Spain.}


\begin{abstract}
Using systematically isothermal coordinates we show that there exist three different maximal extensions of the original Einstein-Rosen bridge. One of them, the hyperbolic Einstein-Rosen bridge, has two-dimensional sections diffeomorphic to the covering space of an hyperboloid of revolution, a singularity satisfying the cosmic censorship and  a bridge  generated by light-like geodesics that can be  traversed by time-like curves. The collapse process that might produce this object is an interesting open problem.\\


\end{abstract}

\maketitle
\section{Introduction}
In $1935$ Einstein and Rosen \cite{ER} proposed the metric
\begin{eqnarray}\label{ER}
&ds^2 =  -\frac{u^2}{u^2+2m}dt^2 + 4(u^2+2m) du^2+(u^2+2m)^2( d\theta^2 + \sin^2 \theta \, d\varphi^2) \\
&-\infty < u < 0 \, ,  0 < u < \infty \ ,
\end{eqnarray}
obtained in two steps, first changing the Schwarzschild radial coordinate $ \rho = u^2 + 2m, \, u > 0$ and then, allowing the new  coordinate to take also values  in $-\infty < u < 0 $, producing two copies of the exterior Schwarzschild metric. But this  metric is degenerate, $ det ( g_{\mu\nu})=0$ at $u=0$, and the absence of good coordinates to cover the union of the two exterior spaces was not considered.\\
 
Recently, Katanaev \cite{KATA1}  obtained this metric as a solution of Einstein's equations with a $\delta$-type energy momentum tensor corresponding to a point particle at $ r= 0$ in isotropic coordinates. 
 But he  pointed out  the incompleteness of its solution, in the same sense as in the original Einstein-Rosen metric, because he did not solve the problem of the coordinate singularity at $u=0$.  Despite that, he demonstrated the complete bridge passability in \cite{KATA2}. 
Guendelman et al.  \cite{GUEN1} claimed that the correct interpretation of the incomplete original Einstein-Rosen bridge has as source  a generalized function (a distribution)  with support on the tridimensional bridge, interpreted as a light-like thin shell (L-L brane). Using Eddington-Finkelstein-like coordinates they obtained an 
 extension with discontinuous first derivatives in the metric, producing the distributional character of the source. The complete passability  of  the bridge, in both senses, and the existence of closed time-like geodesics was discussed in \cite{GUEN2}.\\
 
We must contemplate the notion of extension of  a space-time, and consider 
the possibility of the existence of many different extensions. One can get a maximal extension of a metric, in the sense that all geodesics are complete, but there can exist more than one maximal extension (see for example the nice book of Earman \cite{EARM}).  In fact,   Katanaev commented in \cite{KATA1} that he had to choose  between two possible solutions  for the  lapse function, and that having taken the discarded one he would have obtained the same conclusion as Guendelman et al. \\

The purpose of this work is to study the possibility of other extensions of the Einstein-Rosen space-time. To find them, it will be useful to take into account  that any two-dimensional metric (in our case the metric of the submanifolds $ \theta , \varphi $ constant), verifying some specific conditions, admits isothermal coordinates. So in section \ref{ERBSec} we propose a method for constructing isothermal coordinates, that will be systematically used  to find all possible extensions of the original Einstein-Rosen bridge. We shall obtain first  the  Einstein-Rosen bridge with boundary ($ERb$), which is a non maximal extension of the original Einstein-Rosen metric.  It is also a subspace of the Kruskal-Szekeres space-time \cite{KRUS}, which  has  been  considered as the  maximal extension of the original Einstein-Rosen bridge, but it is not the  only possibility. In section \ref{maximal} we show two other possible extensions by doing a change of topology. One of them, according to Poplawski \cite{POPLA}, should coincide with the aforementioned Guendelman et al.  extension, so we will study the other one. We call it the hyperbolic Einstein-Rosen bridge ($hER$), since it is diffeomorphic to the covering space of an hyperboloid, as we prove in Appendix A. In section \ref{ED} we provide this extension with a differential structure, and obtain its corresponding metric in section \ref{mher}. We find that in this extension the bridge is generated by isotropic geodesics and it is traverable, and furthermore, the first derivatives of the metric are continue on it, in contrast to the Guendelman's extension. However, the metric presents an unexpected singularity in one side, but, as in the Schwarzschild metric, it is concealed by a horizon, which in this case is the bridge itself. Finally, in section \ref{source} we discuss which kind of physical process could generate our extension (or a part of it), and in section \ref{conclusions} we summarize the main conclusions. In order to facilitate the reading, in Appendix B we gather the main definitions used in the paper.

\section{The Einstein-Rosen bridge with boundary $ERb$}\label{ERBSec} 

{\bf{Proposition 1}}: Let us consider a two-dimensional metric  of the  type $ ds^2 = g_{_{11}}(\varphi_2) d \varphi_{_{1}}^2 + g_{_{22}}(\varphi_2) d \varphi_{_2}^2$, with $ g_{11}g_{22} < 0$.
 Under the change of coordinates
\begin{eqnarray}  
&& U = f(\varphi_2) \cosh \varphi_{_1} \, ,   V = f(\varphi_{_2}) \sinh \varphi_{_1}  \, ,\mathrm{ if }\,  \    g_{22} > 0\\
&& U = f(\varphi_2) \sinh \varphi_{_1} \, ,   V = f(\varphi_{_2}) \cosh \varphi_{_1} \, ,\mathrm{ if } \,  \    g_{22} < 0 \ ,
\end{eqnarray}
it  may be written as $ ds^2 = \Omega^2 ( - dV^2 + d U^2)  $ with : $ \Omega^2 = \frac{\mid g_{_{11}}\mid}{f^2} \, ,\  \     f = C e ^{\pm \int \sqrt{\frac{ \mid g_{_{22}}\mid}{\mid g_{_{11}}\mid} }d\varphi_{_2}}$, and $C$ a non null arbitrary constant.\\

The proof is simple. A straightforward calculation transforms the second form of the metric into the first form. 
Let us  apply this result to the metric  $ ds^2 =  -\frac{u^2}{u^2+2m}dt^2 + 4(u^2+2m) du^2$ of the 
$t$-$u$  section of the  Einstein-Rosen bridge.
We must identify in this case  $\varphi_1 = t/k$, taking for $k$ a positive constant to be determined bellow, and  $ \varphi_2 = u$, and substitute $g_{11}= -\frac{k^2 u^2 }{u^2+2m}$, $ g_{22}= 4(u^2+2m)$.
The change of coordinates $ U = f_{_{E R}}(u) \cosh (\frac{t}{\kappa}) \, , \    V = f_{_{E R}}(u) \sinh (\frac{t}{\kappa})$ produces
\begin{eqnarray}
& f_{_{E R}}= C e^{\pm\frac{u^2}{\kappa}}u^{\pm\frac{4m}{\kappa}}\\
& \Omega^2= \frac{\kappa^2}{C^2}\frac{u^2}{u^2+2m} \frac{1}{e^{\pm\frac{2u^2}{\kappa}}u^{\pm\frac{8m}{\kappa}}}
\end{eqnarray}
in the region $ u > 0$, and we can extend it to the region $u<0$. Taking  $\kappa = 4m$ and choosing sign $(+)$ we avoid the metric degeneration, and we get
in the region  $ U^2-V^2 \geq 0 $ the metric  
\begin{eqnarray}
& ds^2 =\frac{16 m^2}{C^2}\frac{e^{-\frac{u^2}{2m}}}{u^2+2m}\left(- d V^2 + dU^2\right)\label{ERI}\\
& U^2-V^2 =  C^2 e^{\frac{u^2}{2m}}u^2\label{E} \ .
\end{eqnarray}
The second equation defines a monotonous  function  $u^2 = h_{ER}(U^2-V^2)$ of the variable $U^2-V^2$, that is derivable in the set $ERb= \{(U,V)\in \mathbb{R}^2 \mid U^2-V^2 \geq 0\} - (0,0)$. 
We will use below  the inverse change given by $ u=\mathrm{sig} (U)\sqrt{h_{ER}(U^2-V^2)}  $.   The set $ U^2 - V^2 = 0 $ is the topological  boundary of the open  $ U^2 - V^2 > 0$, and corresponds to the coordinate axis  $u=0\, (r=2m)$.\\

The constant $C$ may be determined in order to reach the exterior part of the Schwarzschild metric, obtaining $C= 1/\sqrt{2m}$. Using the $h_{ER}$ function defined above we can express the metric in the form
\begin{eqnarray}
&  ds^2 =\Omega^2_{ER} \left(- d V^2 + dU^2\right)\label{ERI1}\\
& \Omega^2_{ER}= \frac{32m^3}{h_{ER}(U^2-V^2)+2m}e^{-\frac{h_{ER}(U^2-V^2)}{2m}}
\label{ERI2}  \ .
\end{eqnarray}
 As in the Kruskal-Szekeres' extension of the Schwarzschild metric (podriem posar la referència de Kruskal \cite{KRUS}), the isothermal coordinates  have solved the trouble at $r=2 m$, but in this case the extension does not cover the region $ U^2-V^2 < 0$. The set of points $ERb= \{(U,V)\in \mathbb{R}^2 \mid U^2-V^2 \geq 0\} - (0,0)$,  is a manifold with boundary  $ \partial ERb = \{ (U,V)\in \mathbb{R}^2  \mid U^2-V^2 =0\}-(0,0)$. 
 The metric is well defined and derivable on $ERb$, and the boundary just corresponds to  the bridge, but it is a bridge to nowhere.
This manifold   is properly an extension of the Einstein-Rosen space, because it adds points to the space that were not covered by the Einstein-Rosen  coordinates. We shall call this extension the Einstein-Rosen bridge with boundary ($ERb$). If we stop here we would have a serious drawback because the radial isotropic geodesics   $ U+V= \mathrm{ const. } \, , \  \  U - V = \mathrm{ const. } $  would be incomplete, as we show in Fig.\ref{ERB}.

\section{Possible maximal extensions}\label{maximal}
\subsection{The Kruskal-Szekeres space-time}\label{KS}
It is trivial to consider the previous extension as a subspace of the Kruskal-Szekeres space-time. In this way  the geodesic incompleteness  at the boundary disappears, but one gets   the singularity at $r=0$. Moreover, only space-like curves might connect the two isometric infinite spaces of the bridge, through the two-dimensional region $\{ U=V=0$ and any angles $\theta, \varphi \}$. Therefore, this extension of the Einstein-Rosen bridge is not traversable.
  Usually, the original Einstein-Rosen metric is extended in this sense \cite{MTW}, but this is not the only possiblity, so we come back to our manifold with boundary to look for alternative extensions. 
\subsection{Two more extensions by changing the topology }\label{free}
\begin{figure}[hbt]\label{ERBcopy}
\centering
\includegraphics[width=0.4\textwidth]{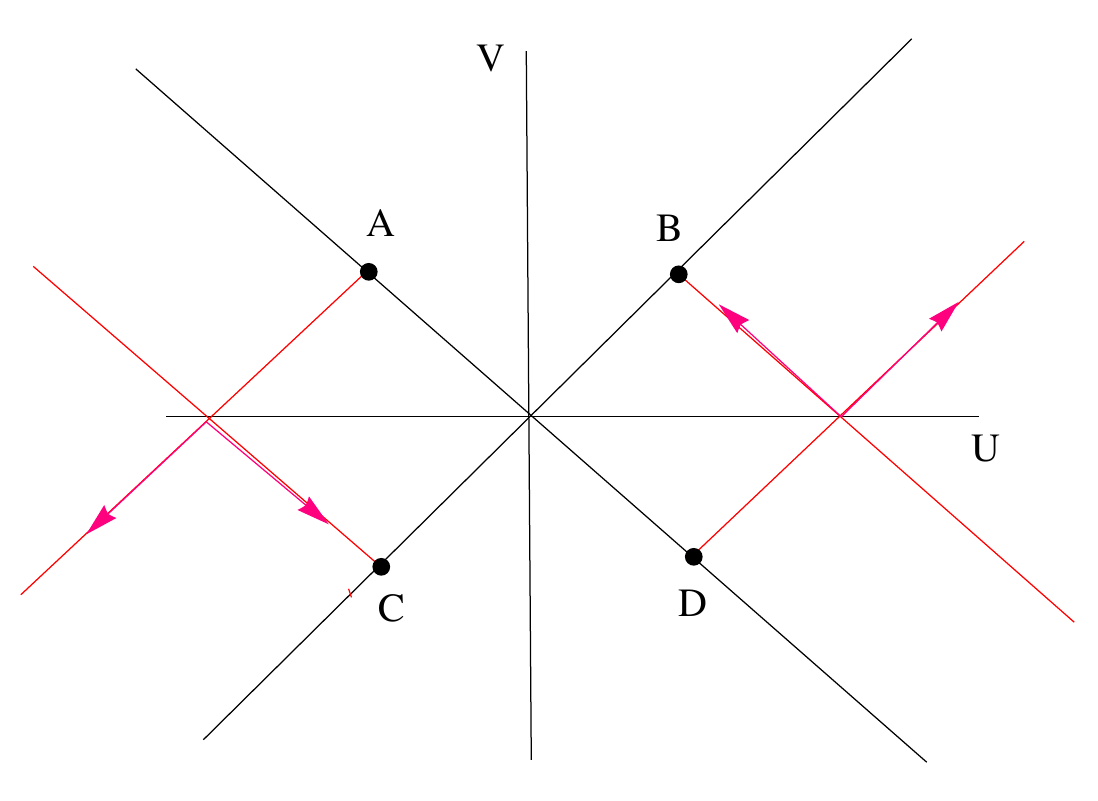}
\caption{\small{Diagram of the $ERb$ extension (where the origin $(0,0)$ is excluded), showing how the incompleteness of the  isotropic geodesics (red lines) may be in principle avoided with a change of topology by gluing points $ A\equiv B , C\equiv D $ , and choosing the appropriated light cones orientations.}}\label{ERB}
\end{figure}

It is well manifest that the metric given in  (\ref{ERI1}) and (\ref{ERI2}) is the same over all the points of the boundary, then we can change the topology by identifying points of it, without come into terms with the field equations. (Let us recall a precedent in the sixties, when W. Rindler \cite{RIND} identified the points $ \{U,V, \theta, \varphi\}$ and $\{ -U,-V, \theta, \varphi\}$ of the Kruskal-Szekeres space-time to construct the Elliptic Kruskal-Schwarzschild space-time. It has the virtue of giving, in the limit $m = 0$, the Minkowski space-time instead of two copies of it, as it happens in the K-S space).
 In principle, as we show in Fig. \ref{ERB}, we have two possibilities:
we can identify points with the same coordinate $V$, i.e; pairs of points as $A$ and  $B$, and pairs as $C$ and $D$, or else we  can identify  points symmetric with respect to the center $(0,0)$, i.e; identify pairs as $A$ and $D$ and pairs as $B$ and $C$. The  change of topology may help to extend the space-time.\\

In the next section {\bf{we shall use the first identification of points of the $ERb$ manifold ($ A\equiv B\, , \, C\equiv D$)}},  and    provide the quotient space $ERb/\hspace{-0.17 cm}\sim$   with a differentiable structure, obtaining what we have called the hyperbolic Einstein-Rosen bridge.
The set formed by all the identified points  defines {\bf{ the bridge $B$ connecting}}  the two sheets  of the Einstein-Rosen space. The other possible identification of points  has been commented by Poplawski \cite{POPLA}
and associated to the extension of the Einstein-Rosen  bridge described by Guendelman et al. \cite{GUEN1}, which has  a $LL$-light brane installed in the  bridge. \\

The following two remarks will help to understand the rest of the paper. Firstly, we  must  realize that   the manifold $ERb/\hspace{-0.19 cm}\sim$  obtained by gluing points will have a different differentiable structure than $ERb$. This means that the   coordinates to be introduced  in a neighbourhood of the bridge $B$ cannot be linked to the coordinates of the $ERb$ manifold of section \ref{ERBSec} by a diffeomorphism. The second remark refers to the orientation of the light cones. Since we are going to identify events $A$ and $B$, if we choose in the region $U > 0$ the orientation down-up for the light cone (as shown in Fig. \ref{ERB}), then  {\bf{ we must choose the  up-down orientation  in the region $U < 0$}}, in order  to make topologically possible that a  light ray can pass through the bridge.

\section{A differentiable structure for the quotient space $ERb/\hspace{-0.17 cm}\sim$}\label{ED}

\begin{figure}[hbt]
\centering
\includegraphics[width=0.4\textwidth]{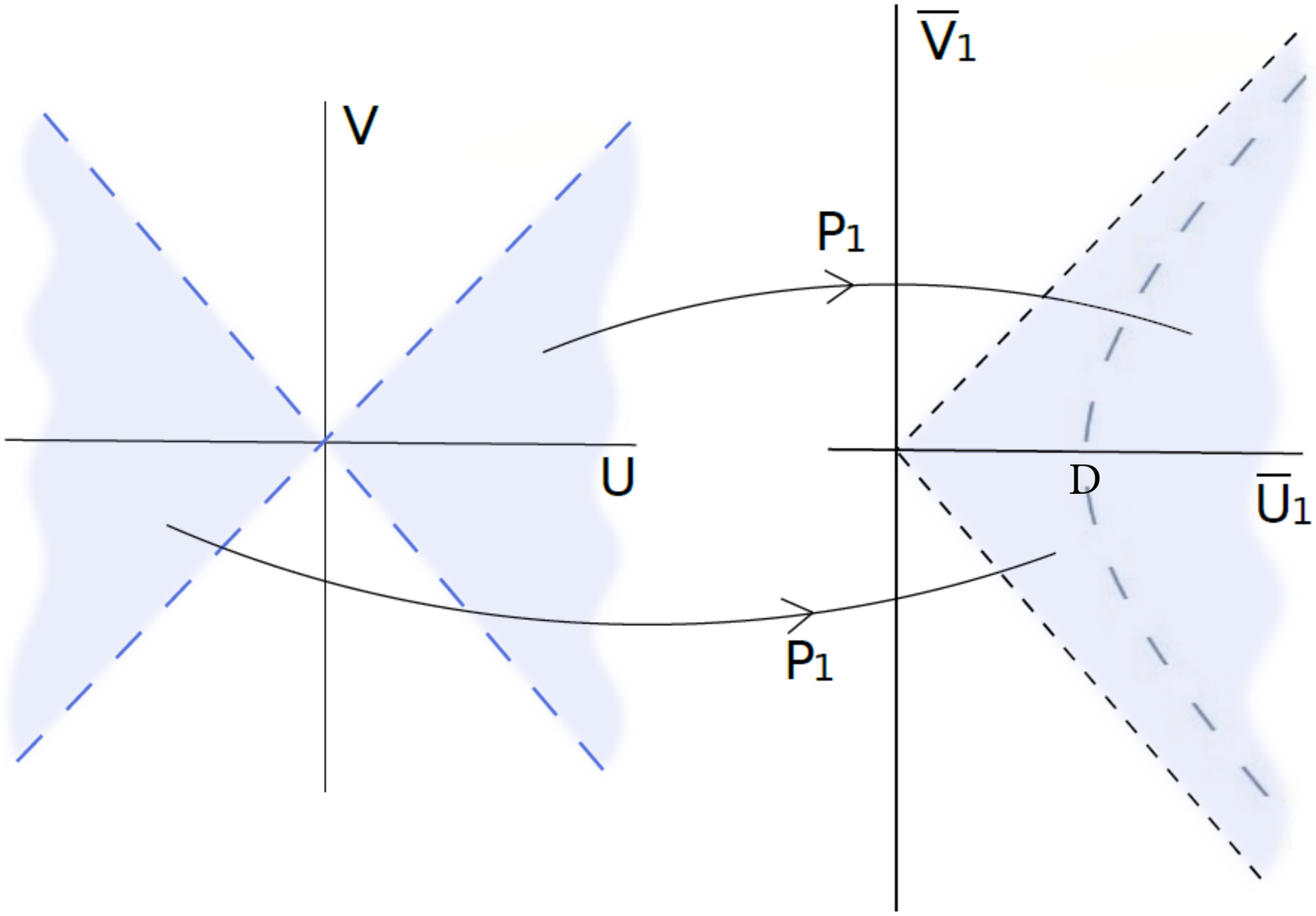}
\caption{\small{Map $P_1: \Gamma_1\rightarrow \bar H_1$. It introduces the coordinates $(\bar U_1 , \bar V_1)$ over $ERb/\hspace{- 0.17 cm}\sim$  excluding the bridge, which is represented as an hyperbola (with point $(D,0)$ removed) in these coordinates.}}\label{cordenadesU1V1}
\end{figure}

\begin{figure}[hbt]
\centering
\includegraphics[width=0.4\textwidth]{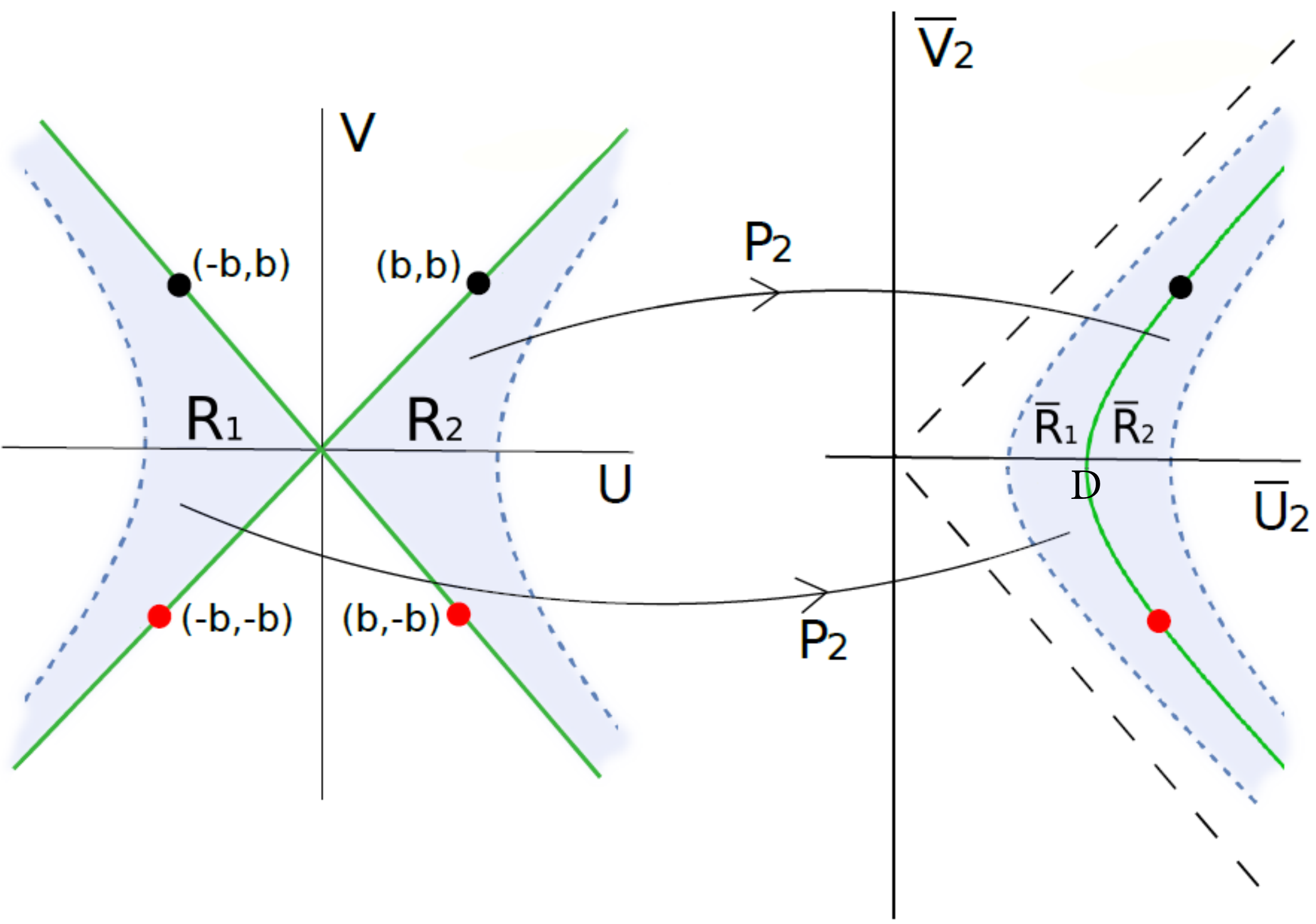}
\caption{ Map $P_2 :  \Gamma_{2}\rightarrow \bar H_2$. It maps   a neighborhood of the bridge $B$, formed by the set of identified points,  onto a neighborhood of the  covering of the throat of the hyperboloid, represented by the hyperbola in green.  The values of this function define the coordinates  $( \bar U_2 , \bar V_2)$.}\label{entorncopy1}
\end{figure}

In order to give a differentiable structure to the quotient space  $ERb/\hspace{-0.17 cm}\sim$ considered in \ref{free},
   we must introduce a system of  compatible coordinates covering all  the space.  We have got it  with only a pair of coordinate systems  that convert  $ERb/\hspace{-0.17 cm}\sim$  into a manifold  diffeomorphic to the covering space of an hyperboloid of revolution  with signature $(- \ +)$. We deviate to the Appendix A the  description of this covering space in  isothermal coordinates, denoted as $(\bar U , \bar V)$, whose values define the set $\bar H = \{(\bar U , \bar V ) \mid  \bar U^2 - \bar V^2 > 0 \, , \bar U > 0\} $ showed in the right hand of Fig. \ref{cordenadesU1V1}-\ref{entorncopy1}. The throat of the hyperboloid generates in the covering space the hyperbola
  $\bar J =\{(\bar U,\bar V)\mid  \bar U^2 - \bar V^2= D^2\, , \bar U > 0\}$, which will correspond to the bridge.  
  We start by introducing a map 
 $P_1: \Gamma_1 \subset ERb \, \rightarrow  \bar H_1\subset \bar H $, defined on the interior of $ERb$:  $\Gamma_1 = \{(U,V)\in \mathbb{R}^2 \mid U^2-V^2 > 0\}$,
and giving values over the covering of the hyperboloid minus the covering of its throat: $ \bar H_1=\{(\bar U_1,\bar V_1)\mid  \bar U_1^2 - \bar V_1^2  >  0 \, , \bar U_1 > 0\} \setminus\bar J$. We define it as follows
\begin{eqnarray}  \label{co1}
 &P_1(U,V)=(\bar U_1 , \bar V_1) \\ 
 &\bar U_1 =  DS\left(\mathrm{sig} (U) \sqrt{h_{ER}(U^2-V^2)}\right)\frac{\mid U \mid}{\sqrt{U^2-V^2}}\\
 &\bar V_1 =  DS\left(\mathrm{sig} (U) \sqrt{h_{ER}(U^2-V^2)}\right)\frac{V}{\sqrt{U^2-V^2}} \ ,
\end{eqnarray}
where $D$ is an arbitrary positive constant and $S(\cdot)$ is a  positive, smooth and monotonous increasing function verifying $S(0)=1$. The values of this function allow the {\bf{introduction of the coordinate system   $(\bar U_1 , \bar V_1)$}} over the subset $\Gamma_1\subset ERb/\hspace{-0.17 cm}\sim$. In Appendix A we prove that, with this definition, the assignation $(\bar U_1,\bar V_1)=(\bar U,\bar V)$ is a diffeomorphism between our manifold minus the bridge and the covering of the hyperboloid minus the throat.

To complete the differentiable  structure we need coordinates for a neighbourhood of the bridge $B$.
To this end, we consider the map $P_2: \Gamma_2 \subset ERb \rightarrow  \bar H_2 \subset \bar H$, with  $\Gamma_{2}= \{(U,V)\mid  0 \leq    U^2 -V^2 < \epsilon^2 \} - (0,0)$, defined as
 \begin{equation} \label{U2V22}
 P_2(U,V)=(\bar U_2 , \bar V_2) = \left( \sqrt{V^2+D^2}+\frac{U^2-V^2}{U}\, \ , V\right) \ .
\end{equation}
The bridge $B$ is mapped onto $\bar J -(D,0)$,  where $\bar J $ is the hyperbola defined above. The image set $\bar H_2= P_2(\Gamma_2)$ is now a neighborhood of the set $\bar J -(D,0)$, defined as
 \begin{eqnarray}
  & \bar H_2 =   \{(\bar U_2,\bar V_2)\mid   \bar U_{L}(\bar V_2) < \bar U_2 < \bar U_{R}(\bar V_2)  \} -(D,0)\label{neighborhood} \\
 & \bar U_{L}(\bar V_2)= \sqrt{\bar V_2^2 + D^2 }-  \frac{\epsilon^2}{\sqrt{\bar V_2^2+\epsilon^2}}\, \ ,  \ 
  \bar U_{R}(\bar V_2)= \sqrt{\bar V_2^2 + D^2 }+ \frac{\epsilon^2}{\sqrt{\bar V_2^2+\epsilon^2}} \ ,
 \end{eqnarray}
  as shown in the right hand of Fig.  \ref{entorncopy1}. This map  verifies $ P_2(-b,b)= P_2(b,b)$ and $ P_2(-b,-b)= P_2(b,-b)$ for any $b >0$, i.e, it is not injective over the boundary $ \partial ERb$ defined in \ref{ERBSec}.  The set $\Gamma_2 \setminus \partial ERb $ is formed by two disjoint open subsets $ \{R_i\}_{ i= 1,2}  \subset \Gamma_2 \setminus \partial ERb$, and $P_2$ maps them bijectively onto two open subsets $ \{\bar R_i\}_{ i= 1,2}  \subset  \bar H_2\setminus\ \bar J $, i.e,   $P_{2}( R_i) = \bar R_i$ as shown in Fig. \ref{entorncopy1}. 
 We shall need the coordinates of the sets $R_i = P_{2}^{-1}( \bar R_i)$ as functions of the coordinates of $\bar R_i$.   One gets  
 \begin{eqnarray}
 && P_{2}^{-1}(\bar R_1)= \{( U_{-}, V)\}\, \ \ , \, \ \  P_{2}^{-1}(\bar R_2)= \{(U_{+},V)\}\label{pedos}\\
 && U_{+} = \frac{1}{2}(\sqrt{\varphi^2 +4 \bar V_2^2} - \varphi ) \,  \   , V=\bar V_2 \label{antecedent+}\\
&& U_{-} = -\frac{1}{2}(\sqrt{\varphi^2 +4 \bar V_2^2} + \varphi ) \,  \   , V=\bar V_2 \, \ , \ \varphi = \sqrt{\bar V_2^2+ D^2}-\bar U_2\label{antecedent-} \ .
 \end{eqnarray}
The non-injectivity of $P_2$ over the boundary $ \partial ERb$
 will make possible  to assign the same coordinates to pairs of points like  $A , B $ and  $C , D$. By identifying the pairs  $(-b,b) , (b,b)$ and  $(-b,-b) , (b,-b)$ of the set $\Gamma_2, 
$ shown in the left part of Fig. \ref{entorncopy1}, we obtain a neighborhood of the bridge of the quotient space $\mathrm{ERb} /\hspace{-0.17 cm}\sim$.\\

With this application we can proceed to the {\bf{introduction of the coordinate system $(\bar U_2 , \bar V_2)$}}
in a  neighborhood $\Gamma_2/\hspace{-0.17 cm}\sim \, \ \subset ERb /\hspace{-0.17 cm}\sim$  of the bridge of  the quotient space  as follows. To the pair of  points $ (-V,V) \equiv (V,V)$ we assign  coordinates $(\bar U_2 , \bar V_2) = ( \sqrt{ V^2+ D^2}, V)$. To  points such that  $U^2 - V^2 > 0$ we assign coordinates  $(\bar U_2 , \bar V_2) = (\sqrt{V^2 + D^2} +\frac{U^2-V^2} {U}, V)$. With the next proposition we shall provide a differentiable structure to the quotient space  $ERb /\hspace{-0.17 cm}\sim$. \\
  
{\bf{Proposition 2}}. The  coordinate systems   $(\bar U_1 , \bar V_1) $ and $(\bar U_2 , \bar V_2) $ over the quotient space $ERb /\hspace{-0.17 cm}\sim$ are compatible, i.e., the function change of coordinates is a diffeomorphism.\\

\begin{figure}[hbt]
\centering
\includegraphics[width=0.5\textwidth]{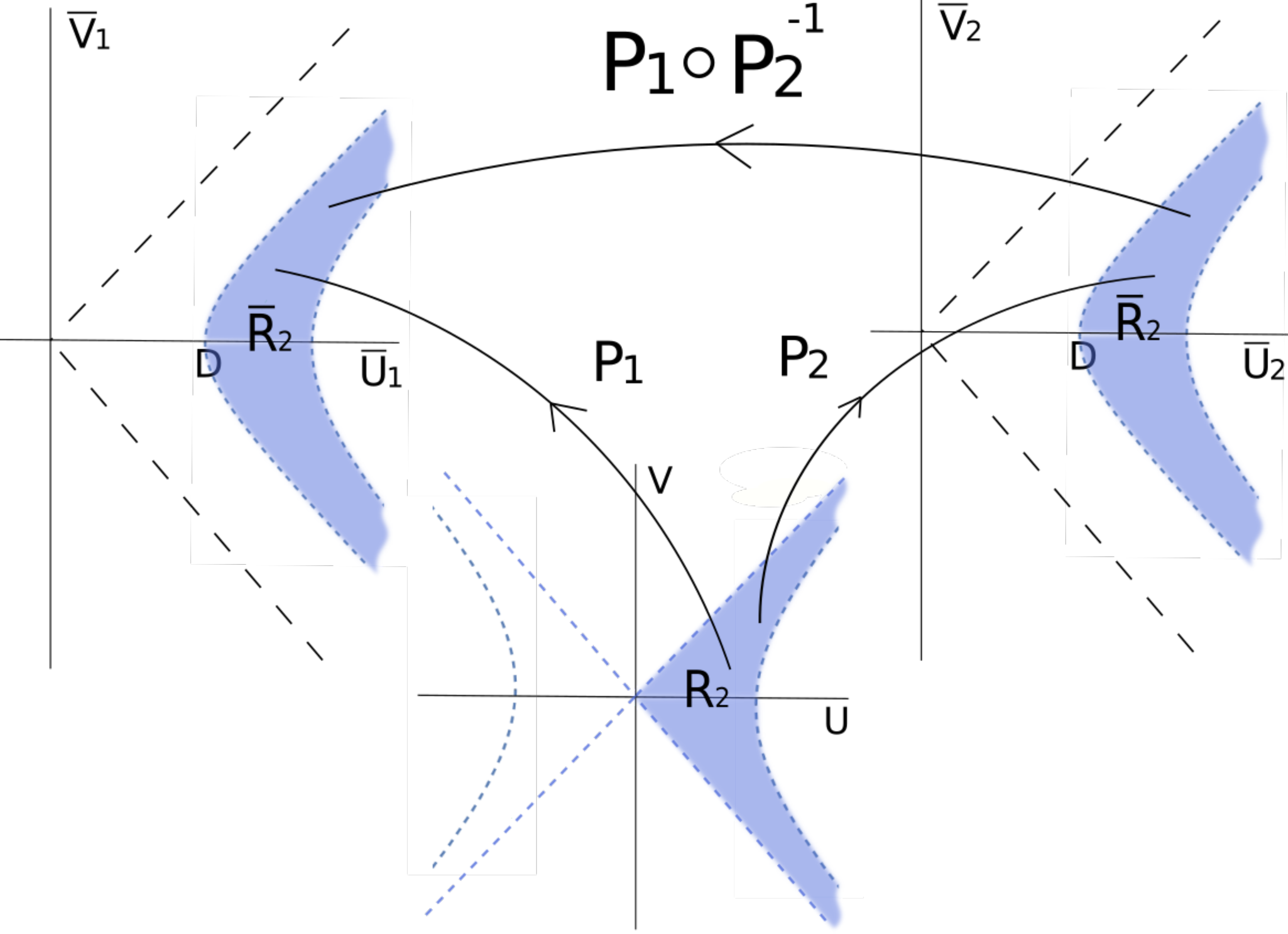}
\caption{Diagram of the change of isothermic  coordinates  $ ( \bar U_2 , \bar V_2) \rightarrow (\bar U_1 , \bar V_1)$ in the region $R_2$, that has been proved to be differentiable.}\label{entorncopy}
\end{figure}

To prove it we have to consider the common regions of the domains of $P_1$ and $P_2$, that is $R_1\cup R_2$. It is enough to prove it in one of them, since in the other one the procedure would be analogue. We choose $R_2$, as shown in Fig. \ref{entorncopy}. So, we shall choose  $U =U_{+}=  \frac{1}{2}(\sqrt{\varphi^2 +4 \bar V_2^2} - \varphi ), \bar V_2 = V$ given in (\ref{antecedent+}), and
we can express $V/U$ and $U^2-V^2$ as functions of $(\bar U_2 , \bar V_2)$
\begin{eqnarray}
& \frac{V}{U}= q ( \bar U_2,\bar V_2) \, \ , \  q ( \bar U_2,\bar V_2)= \frac{2\bar V_2}{- \varphi +\sqrt{\varphi^2  + 4 \bar V_2^2} }\\
& U^2 -V^2 = p ( \bar U_2,\bar V_2)\, \ , \  p ( \bar U_2,\bar V_2)= \frac{\varphi}{2}\left(\varphi- \sqrt{\varphi^2 + 4\bar V_2^2 } \right)\label{u2mev2} \ .
\end{eqnarray}
The functions $p ( \bar U_2,\bar V_2)$  and $q ( \bar U_2,\bar V_2)$ are both  differentiable and
 moreover $|q ( \bar U_2,\bar V_2)| < 1$  because $\varphi < 0$ in the region considered. This property will be used bellow. Let us introduce $ \tilde S(\bar U_2,\bar V_2)= S\left(\sqrt{h_{ER}(p(\bar U_2,\bar V_2)}\right)$, then from equation (\ref{co1})  we have
\begin{eqnarray}
& \bar U_1^2 -\bar V_1^2 = D^2 \tilde S(\bar U_2, \bar V_2)^2\label{una}\\
&  \frac{\bar V_1}{\bar U_1}= q( \bar U_2,\bar V_2)= \frac{V}{U}\label{dues}  \ ,
\end{eqnarray}
 and by  combining (\ref{una}) and (\ref{dues})  one gets the change of coordinates 
\begin{equation}\label{1-2}
 \bar U_1 =\frac{D \tilde{S}(\bar U_2 , \bar V_2) }{ \sqrt{1-q ( \bar U_2,\bar V_2)^2}} \, \ , \  \bar V_1 =\frac{D  \tilde{S}(\bar U_2 , \bar V_2)  q (\bar U_2,\bar V_2)}{\sqrt{1-q ( \bar U_2,\bar V_2)^2}}  \ .
\end{equation}
 The change of coordinates (\ref{1-2}) is differentiable because $ q^2 < 1$ in the domain considered. This completes the proof. \\

With the  coordinate  systems $ (\bar U_1 , \bar V_1) , (\bar U_2 , \bar V_2) $ we cover all  the space $\bar H = \{(\bar U , \bar V ) \mid  \bar U^2 - \bar V^2 > 0 \, , \bar U > 0\} $. 
  {\bf{Hence  we have given a differentiable structure to the quotient space}} $ERb /\hspace{-0.17 cm}\sim$.  Regarding the geometrical characteristics of this space, the manifold  $ ERb/ \hspace{-0.17 cm}\sim$ 
  is diffeomorphic to the covering space of the hyperboloid of revolution with the point $(D,0)$ removed, as proved in Appendix A. This is  why we have given the name \textbf{hyperbolic Einstein-Rosen bridge (\textit{hER})} to the manifold  $ ERb / \hspace{-0.17 cm}\sim$. Simply, to a point with coordinates $(\bar U_1, \bar V_1)$ we associate a point of the hyperboloid with the same coordinates $ (\bar U , \bar V)=(\bar U_1, \bar V_1)$, and similarly for points of $ERb/\hspace{-0.17 cm}\sim$  with coordinates $(\bar U_2, \bar V_2)$. Therefore, the set $\bar J - (D,0) \subset \bar H$, which is the throat of the hyperboloid, is diffeomorphic to the bridge $B \subset ERb/ \hspace{-0.17 cm}\sim$ that connects the sheets of the Einstein-Rosen space. \\

In the next section, using the function $P_{2}^{-1}$ introduced in  (\ref{pedos}) we shall get  the metric  in a neighborhood of the bridge expressed in coordinates $(\bar U_2 , \bar V_2)$, and will be able to study the passability of the bridge.

\section{The  metric on the  hyperbolic Einstein-Rosen bridge $hER$}\label{mher}

   The metric space $(ERb, \Omega^2_{ER} \eta) $ with $ \eta= - dV^2 + dU^2 $, obtained in section \ref{ERBSec}, is a non maximal extension of the incomplete Einstein-Rosen bridge. In the previous section we have obtained, by identifying points of the bridge, a new manifold $ERb/\hspace{-0.17 cm}\sim$  which is diffeomorphic to the covering space of an hyperboloid of revolution, and we have provided it with a differential structure. We shall explain how to obtain in this new space a maximal extension of the Einstein-Rosen bridge different from the well-known Kruskal-Szequeres black hole. \\
   
The dipheomorphism $P_{2}^{-1}: \bar R_2 \subset \bar H_2 \setminus \bar J \rightarrow  R_2 \subset \Gamma_2$   will be used now to pullback the metric  on the open set $R_2\subset\Gamma_2$ over the open set $\bar R_2$. 
  So, taking into account (\ref{antecedent+}), we get
\begin{eqnarray}
& d  U = A d \bar U_2 + B d  \bar V_2 \,  , \,  d V  = d \bar V_{2} \\
&  A = \frac12\left(1-\frac{ \varphi}{\sqrt{\varphi^2 + 4\bar V_2^2}}\right)\label{A}\\
&  B= - \frac{1}{2}\left(1-\frac{ \varphi}{\sqrt{\varphi^2 + 4\bar V_2^2}}\right)\frac{\bar V_2}{\sqrt{\bar V_2^2 + D^2}} +\frac{ 2 \bar V_2}{\sqrt{\varphi^2 + 4 \bar V_2^2}}\label{B} \ ,
\end{eqnarray}
with  $\varphi = \sqrt{\bar V_2^2+ D^2}-\bar U_2 $. Using these relations we get the metric in the region $\bar R_2 \subset \bar H_2$, where $\varphi < 0$,  shown in  Fig.  \ref{entorncopy1}
 \begin{eqnarray}\label{metric2}
 &d\bar s^2 = \bar \Omega^2(\bar \eta_{_{\bar U_2 \bar U_2}} d \bar U_2^2+ \bar \eta_{_{\bar V_2 \bar V_2}} d\bar V_2^2 +2 \bar \eta_{_{\bar U_2 \bar V_2}}d \bar U_2 d \bar V_2)
\\
&\bar \eta _{_{\bar V_2 \bar V_2 }}= B^2 -1 \, \ , \ \bar \eta_{_{\bar U_2 \bar U_2} }=A^2 \, \ , \, \bar \eta_{_{\bar U_2 \bar V_2} } = AB \ ,
\end{eqnarray}
where the factor $\bar \Omega^2=\Omega^2_{ER}(U^2-V^2) $ is obtained by substituting $U^2 - V^2 =U_+^2 - V^2= \frac{\varphi}{2}\left(\varphi- \sqrt{\varphi^2 + 4\bar V_2^2 } \right)$into the function $\Omega^2_{ER}$ given in  (\ref{ERI2}). 
 The same functional form of the metric defined above in $\bar R_2$ {\bf{is extended}} now to the the region $ (\bar J -(D,0)) \cup \bar R_1$ where $\varphi \geq 0$. 
The extended metric is degenerate over the axis $\bar V_2=0$ in the region $\varphi > 0$ (the left hand of the bridge), because there we have $ A=B=0$, but it  is smooth over all the bridge $\bar J -(D,0)$.\\

The components of the metric are analytic functions in the open set formed by subtracting the semi axis $\varphi > 0 \, , \bar V_2 =0$ from the open set $\bar H_2$, and so are the components of the Ricci tensor. On the other hand, by construction, the Ricci  tensor is null in the region $\bar R_2$ (to the right of the bridge), since this region is diffeomorphic to the right hand of the Einstein-Rosen space-time, where the Ricci tensor vanishes. Therefore, according to the theorem of interior uniqueness of analytic functions, the Ricci tensor will also be null in the region $\bar R_1$ (to the left of the bridge) minus the semi axis $\varphi > 0 \, , \bar V_2 =0$. Then, the metric verifies the Einstein equations in empty space, except in the degenerate region. To ascertain the geometrical character of the degeneracy  we have considered the Kretschmann scalar, $\mathrm{Kret}(\bar U_2, \bar V_2)=R_{abcd}R^{abcd}$ in the region $\varphi > 0$, $\bar U_2 < D$, to study the limit when $\bar V_2$  tends to zero keeping $\bar U_2$ constant.  We have found that it diverges as $ (D-\bar U_2)^6/\bar V_2^{10} $. The calculation is straightforward and helped by the fact that the metric is the sum of two mutually independent submetrics.
 The semi-axis  $ \varphi > 0 , \bar V_2 = 0 $  is therefore a geometrical singularity of the space-time. \\

 We shall show that the bridge is a horizon that protects the right region ($\varphi < 0$) from the singularity present in the left region ($\varphi > 0$), as in the maximal extension of the Schwarzschild metric.
We start  studying the geometrical properties of the hyperbola $\bar J -(D,0)$.
A bridge is the possibility of communicating two separate zones. To ascertain if our hyperbola has this property  we shall construct the light cones over it, i.e., the tangent vectors to the pair of light rays at any point of $\bar J$.
If we renounce to an affine parametrization we can  get  them   by considering the simpler metric $ ds^2 =\bar \eta_{_{\bar U_2 \bar U_2}} d \bar U_2^2+ \bar \eta_{_{\bar V_2 \bar V_2}} d\bar V_2^2 +2 \bar \eta_{_{\bar U_2 \bar V_2}}d \bar U_2 d \bar V_2$.  It is manifest that the transformation $ \bar V_2 \rightarrow - \bar V_2$ is an isometry, therefore the geodesics in the $\bar V_2 < 0$ region may be obtained by symmetry respect the $\bar V_2=0$ axis.
Using the coordinate $\bar V_2$ as parameter of the isotropic geodesics, we get an algebraic equation to determine the isotropic directions, which is given by
\begin{equation}
  \bar \eta_{_{\bar U_2 \bar U_2}}\left( \frac{d \bar U_2}{d \bar V_2}\right)^2 + 2 \eta_{_{\bar U_2 \bar V_2}} \frac{d \bar U_2}{d \bar V_2} + \bar \eta_{_{\bar V_2 \bar V_2}} =0  \ .
\end{equation}
The  two solutions  define the following two first order differential equations
\begin{equation}
\frac{d \bar U_2}{d \bar V_2}= \frac{1-B}{A}\, \  \ ,  \  \   \frac{d \bar U_2}{d \bar V_2}=-\frac{1+B}{A}\label{isoge12}  \ ,
\end{equation}
where $A$ and $B$ are defined in (\ref{A}), (\ref{B}). On the bridge it is verified $\varphi =0$, so we have 
\begin{eqnarray}
&& \frac{1-B}{A}\mid_{\varphi =0} = \frac{\bar V_2}{\sqrt{\bar V_2^2 + D^2}} \, \  \ , \    -\frac{1+B}{A}\mid_{\varphi =0} =  \frac{\bar V_2}{\sqrt{\bar V_2^2 + D^2}}-4 \,  \     \mathrm{if}\,  \  \bar V_2 > 0\\
&& \frac{1-B}{A}\mid_{\varphi =0} = \frac{\bar V_2}{\sqrt{\bar V_2^2 + D^2}}+4 \, \  \ , \    -\frac{1+B}{A}\mid_{\varphi =0} = \frac{\bar V_2}{\sqrt{\bar V_2^2 + D^2}} \,  \     \mathrm{if}\,  \  \bar V_2 < 0 \ ,
 \end{eqnarray}
and it follows immediately that the parametrization  $( \bar U_2 = \sqrt{\bar V_2^2 + D^2}, \bar V_2 = \bar V_2)$ of the  bridge $\bar J-(D,0)$   is a light ray. One of the two isotropic vectors at any point  $ p \in \bar J-(D,0)$ is the tangent vector to the bridge $K_1= (\frac{\bar V_2}{\sqrt{\bar V_2^2 + D^2}},1 )$,  while  the other one is clearly directed from right to left:  $K_2 =( -4 + \frac{\bar V_2}{\sqrt{\bar V_2^2 + D^2}} , 1) $ in the $\bar V_2 > 0 $ region, and from left to right $K_2 =( 4 + \frac{\bar V_2}{\sqrt{\bar V_2^2 + D^2}} , 1) $ in the $\bar V_2 < 0 $ zone . The couple $(K_1,K_2) $ define the light cone at each point of the bridge.  They show that  in the region $\bar V_2 > 0$  the matter can traverse the bridge only in right to left direction; by contrary  in the region $\bar V_2 <  0$  matter only can traverse it from left to right. Therefore we conclude that the bridge is traversable.\\

\begin{figure}[hbt] \label{consher}
\centering
\includegraphics[width=0.35\textwidth]{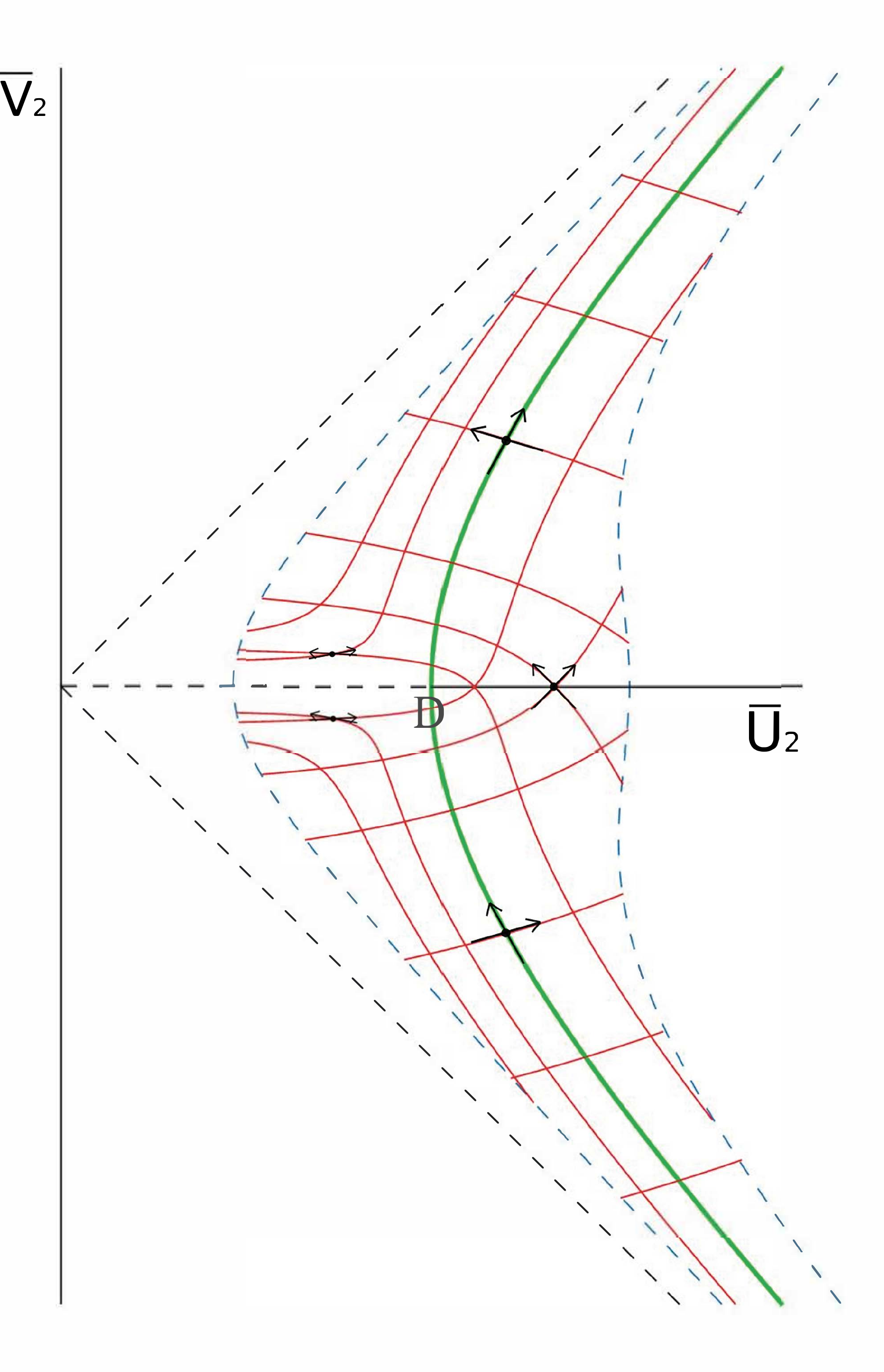}
\caption{Diagram showing the light rays (red lines) in  a neighbourhood of the bridge using   coordinates $(\bar U_2, \bar V_2) $. The central hyperbola (green line), minus $(D,0)$,  is the bridge formed by two interrupted  light rays. The light cones traced on the bridge show the possible directions of transit for light and matter. All  the light  rays have been obtained using  $Mathematica$.}\label{Fi:condellum1.pdf}. 
\end{figure}

Fig. \ref{Fi:condellum1.pdf} shows the representation of some isotropic geodesics in a neighbourhood of the bridge and the light cones, which show the possible transit directions. These directions agree with the orientation of the  light cone  chosen  at the end of  section \ref{free}. Let us point out that if we  had started  obtaining the metric on $\bar R_1\subset \bar H_2$ by choosing the value $U = U_{-}$ given in  (\ref{antecedent-}), instead of $ U = U_{+}$, and extending it to $\bar R_2$, the bridge would be   passable too but with inverted transit directions. This option would correspond to the opposite choice of the light cone orientation in  section (\ref{free}).\\

Regarding the point $(D,0)$, it is a conical singularity and is not included in the space-time. If it had been included in the new manifold, the metric would be discontinuous at that point, and equations (\ref{isoge12}) would be of the form $\frac{d\bar U_2}{d\bar V_2}= f(\bar U_2,\bar V_2)$ with $f(\bar U_2,\bar V_2)$ discontinuous at $(D,0)$, no satisfying the existence theorem. \\

  An interesting characteristic of  the semi-axis  $( \varphi > 0 , \bar V_2 = 0 )$, where the metric is degenerate, is that it cannot be connected by light rays to any point of the space time, as it  is manifest in Fig. \ref{Fi:condellum1.pdf}. Moreover, light and matter coming from the region $\varphi > 0 , \bar V_2 > 0$ cannot traverse to the right side. Therefore, the bridge acts as a horizon and the singularity satisfies the cosmic censorship, since it is causally disconnected from the right side. This singularity could have been avoided by obtaining the metric on the left  and right regions (i.e., $\bar R_1$ and $\bar R_2$) independently, as  pull-backs  of the known metrics in regions $ R_1$ and $R_2$. But the metric obtained in this way is not continuous on the bridge, and it cannot be considered an actual extension of the Einstein-Rosen space.\\

\section{The problem of the source of the $hER$ metric space.} \label{source}

 Finally, let us add some comments about  the source of the $hER$ metric space.  According to Katanaev,  a point particle is the  source of the incomplete  metric (\ref{ER}). However,  the point particle is at $r=0$ in isotropic coordinates, which is at  infinite distance of the bridge
where the curvature of this space-time is enormous. This fact is interpreted by Katanaev as repulsive gravity in the left side of the bridge, 
 but for us this is  an uncomfortable conclusion that prompts to look for a different interpretation.\\
 
\begin{figure}[hbt]
\centering
\includegraphics[width=0.35\textwidth]{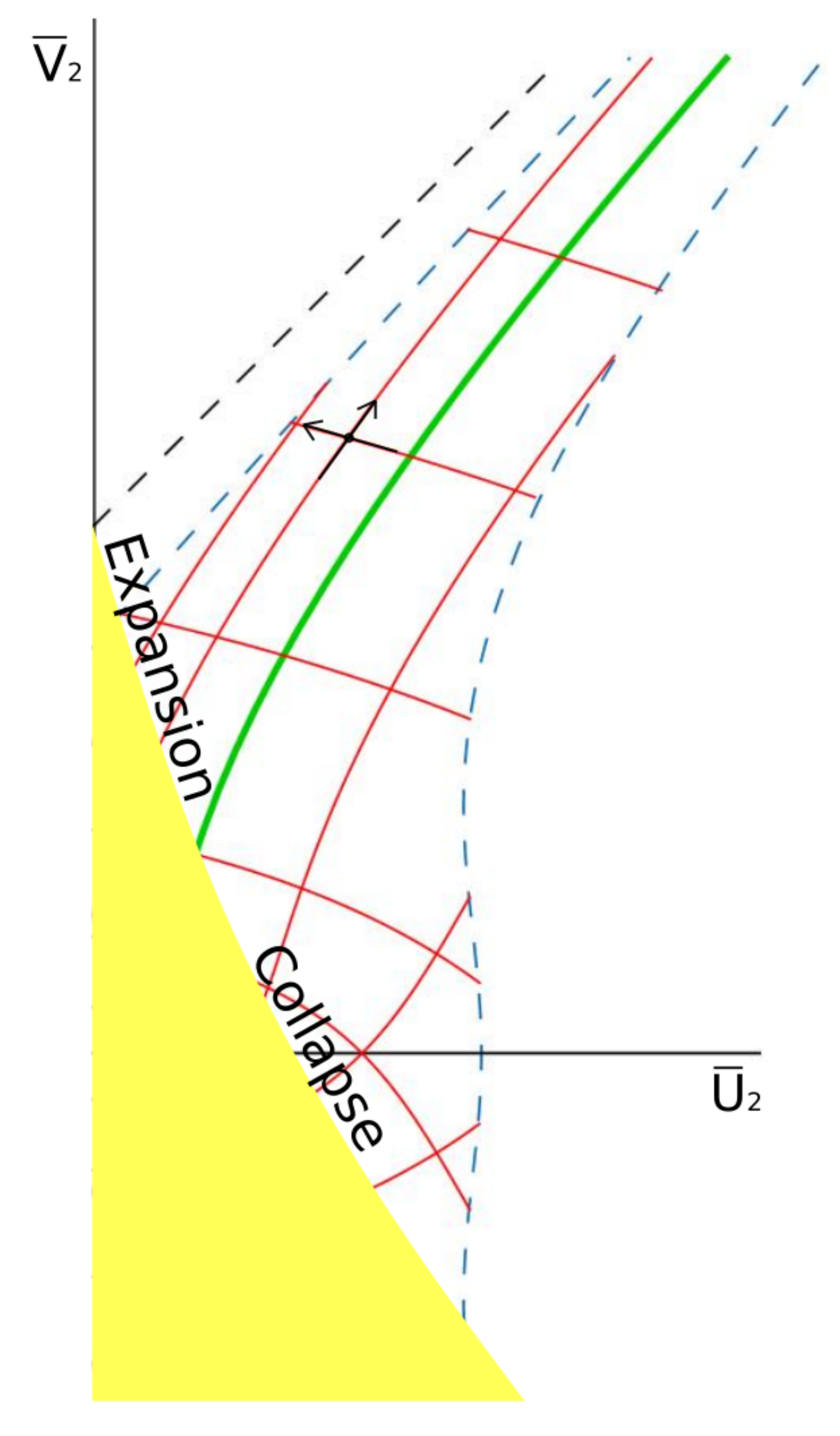}
\caption{ Diagram of the proposed collapse process generating a part of the $hER$ space-time. The green line represents the bridge, generated by light rays.   The yellow  region is the source, contracting before and in expansion after passing the bridge.}\label{colapse}
\end{figure}

In the case of the black-hole extension of the Schwarzschild metric,  an illuminating finding was its relation with the {\bf{continued gravitational contraction}} of an  spherical star made of pressure-less  matter \cite{OPENSNY}: only the exterior of the collapsing star was  a part of the Kruskal-Szekeres extension. In our case, it is an interesting  issue  to find the kind of collapse whose exterior would correspond to a part of the hyperbolic  Einstein-Rosen bridge. The spherical star must contract, as shown in Fig. \ref{colapse}, until the free surface reaches the bridge, and then it must begin to expand again, but now in the other side of it. (Only  negative pressures could produce this sort of collapse. Newtonian gravity predicts a negative gravitational contribution to the pressure, but a correct extension of this gravitational effect has not been accomplished   so far). The orientation of the light cone in Fig. \ref{colapse} shows that  the source in her expansive phase, in the left part of the bridge, can not be observed from the right of it, and no material particle evolving in the left part can get the bridge.  In other words, an observer in the right side could not distinguish this process from an Oppenheimer-Snyder collapse creating a black-hole. However, after crossing the horizon, the star would not finish in a singularity, but would begin to expand again. In fact, no singularities would be present, since the degenerate region $(\bar U_2 < D , \bar V_2=0)$ is now occupied by the collapsing fluid.\\

This interpretation would explain why gravity is stronger  on the bridge, because any point of it is  connected by a light ray (the  bridge) to the collapsing  body at the point of maximum contraction. By contrast, the curvature tends to zero  at great distances to the left of the bridge, because  the light rays connect with points at the expansion phase. 
 No point-particle would be present in this space-time, 
  only a part of the mathematical point-particle  Katanaev's  solution  would be  generated as the exterior of this collapse-expansion process.  \\

\section{Conclusions} \label{conclusions}

 In this work we have obtained a new extension ($hER$) of the original incomplete Einstein-Rosen bridge, different from the well-known Kruskal-Szekeres space-time and from the more recent one obtained by Guendelman et al. \cite{GUEN1,GUEN2}. The manifold of $hER$ is diffeomorphic to the covering space of an hyperboloid of revolution, and the metric components are analytic functions in a neighborhood of the bridge with a geometrical singularity on the semi axis $\bar V_2 = 0 , \bar U_2 < D$. The bridge is formed by light rays, and is traversable unlike the bridge in the $KS$ extension. The $hER$ space-time differs from the one obtained by Guendelman et al. in three aspects: a) it has no source installed in the bridge, b) it has no closed time-like geodesics typical of whormholes with exotic sources, and c) it presents a singularity, though compatible with the cosmic censorship conjecture (it is causally disconnected from the opposite side of the bridge, so the bridge acts as a horitzon). Finally, we have discussed the problem of the source of this space-time, and proposed a gravitational collapse as in the black-hole case but with two phases: contraction in one side and expansion in the other. It would generate a part of the $hER$ metric in the exterior of the collapse, avoiding the singularity. We leave the deep analysis of this process for future work.\\

{\it Acknowledgments.--} P. B. is supported by a Ph.D. fellowship, Grant No.  FPU17/03712. M.P. thanks support by the Spanish "Ministerio de Economia y Competitividad (sic)" and the "Fondo Europeo de Desarrollo Regional" MINECO-FEDER Project No. PGC2018-095251-B.100.

\section*{Appendix A: Diffeomorphism with the covering space of the hyperboloid}\label{HER}
We shall obtain the expression in isothermal coordinates of the covering space of a two dimensional hyperboloid with signature $(-,+)$ that has been used in section \ref{ED}. 
The manifold  of the  two-dimensional sections with ($\theta , \varphi$) constant  of the Einstein-Rosen metric given in (\ref{ER}) could be considered, excluding the section $u=0$, as  a surface of revolution (for example the one sheet hyperboloid) if we take $t$ as the angle of revolution. This is only possible if we consider the covering space of the hyperboloid, $\bar H,$ by extending  to all $\mathbb{R}$ its angular coordinate, defined in $(0, 2\pi)$.  We shall do that  and   our quotient space  $ERb/\hspace{-0.17 cm}\sim$ defined in section \ref{ED} will be   
 diffeomorphic to the covering  surface $\bar H$.\\
 
 {\bf{Description of a one sheet hyperboloid with signature $ ( - \, \ +)$}}. We consider in the space  $\mathbb{R}^3$ the metric $ ds^2 = -dx^2 -dy^2 +dz^2 $, and the surface of an hyperboloid defined  by  $ \frac{x^2}{q^2} +\frac{y^2}{q^2}-\frac{z^2}{p^2}=1$, with $ p > q $. It can be parametrizated as: $X(\varphi_1 , \varphi_2)= ( q \cosh \varphi_2  \cos \varphi_1, q \cosh \varphi_2  \sin \varphi_1 , p \sinh  \varphi_2  ) $,  with $ 0  < \varphi_1 < 2\pi \, , \  - \infty < \varphi_2 < \infty $, and one obtains  the  metric  $ds^2 = -q^2 \cosh^2\varphi_2 \,d  \varphi_1^2 + ( p^2 \cosh^2\varphi_2 - q^2 \sinh^2\varphi_2 ) d \varphi_2^2$. Now, using  the
  {\bf{Proposition 1}} we  can introduce  isothermal coordinates 
  $ \bar U = f_{{H}}(\varphi_2) \cosh (\varphi_1)$, 
   $ \bar V = f_{{H}}(\varphi_2) \sinh (\varphi_1)$, 
   with $ f_{H}(\varphi_2)= D \exp(\int_0^{\varphi_2} \sqrt{1-\frac{p^2}{q^2}  \tanh^2s}\,  ds )$ and $D$ a  positive arbitrary constant, and  express the  metric as $d\bar s^2 = q^2 \cosh^2 (\varphi_2) f_{H}^{-2}(\varphi_2) ( -d \bar V^2 + d\bar U^2)$.
 The equation $ \bar U^2 -\bar V^2 =  f_{H}^2(\varphi_2)$ defines  a monotonous increasing  function,  $ \varphi_2 = h_{_{H}}(\bar U^2-\bar V^2)$ in the open set  $\bar U^2-\bar V^2>0 , \bar U>0$. It  verifies: $  h_{H}(0)=-\infty \, , \  h_{H}(D^2)= 0 \, , \ h_{H}(\infty)=\infty $.  In isothermal coordinates, 
 the lines $\varphi_1 = \mathrm{const}.$ in the $( \varphi_1 , \varphi_2)$ plane  transform into the straight lines $ \bar V = \tanh(\varphi_1)\, \bar U$ in the $(\bar U , \bar V)$ plane, for values $0 < \varphi_1 < 2 \pi$; 
  and the circles  $\varphi_2 = \mathrm{const}.$ into segments of  the hyperbolae $\bar U^2 -\bar V^2 = f_H^2(\varphi_2)$.  In particular, the circle  $\varphi_2 = 0$ transforms into a segment of
the hiperbola $\bar J= \{(\bar U,\bar V)\mid \bar U^2 -\bar V^2 = D^2, \bar U>0 \}$ in isothermal coordinates. \\

  {\bf{The covering space $\bar H$ of the  one sheet hyperboloid}}. 
The covering space of a circle in $ \mathbb{R}^2$, parametrized as $ X(\varphi_1,\varphi_2)=( a \cos \varphi_1, a \sin \varphi_1)$,  is the helix in $ \mathbb{R}^3$, parametrized as $X(\varphi_1,\varphi_2)=( a \cos \varphi_1, a \sin \varphi_1 , b \varphi_1)$. Using this analogy we extend  the coordinate $\varphi_1$ to obtain the parametrization of the covering space of the one sheet hyperboloid:
$ X(\varphi_1 , \varphi_2)= ( q \cosh \varphi_2  \cos \varphi_1, q \cosh \varphi_2  \sin \varphi_1 , p \sinh  \varphi_2  , \  \varphi_1 )$, with $- \infty < \varphi_1 < \infty\, , \  - \infty < \varphi_2 < \infty$.
 The expression in isothermal coordinates of this covering space can be obtained by extending  the range of values of the isothermal coordinates of the hyperboloid to the open set $\bar H= \bar U^2 - \bar V^2 > 0\, , \bar U > 0$. \\


{\bf{Diffeomorphism $ERb/\hspace{-0.17 cm}\sim\, \leftrightarrow \bar H-(D,0)$}}. We start  with the non isothermal coordinates $ ( u , t ), (\varphi_2 , \varphi_1)$, used to express  the Einstein-Rosen metric (\ref{ER}) and the covering space of the one sheet hyperboloid respectively. Let us consider the following diffeomorphisms: $\varphi_2 =  u\,,\,\varphi_1 = \frac{t}{4m}$, for the open $u>0$ and $\varphi_2 =  u\,,\,\varphi_1 = -\frac{t}{4m}$ for the open $u<0$. Now, by considering the inverse change $ u=\mathrm{sig} (U)\sqrt{h_{ER}(U^2-V^2)}$ (see section \ref{ERBSec}) and the well-known  hyperbolic trigonometric relations, we  get the expressions   of this assignations in isothermal coordinates
\begin{eqnarray}
&  \bar U =  f_H\left(\mathrm{sig}(U)\sqrt{h_{ER}(U^2-V^2)}\right)\frac{\mid U \mid}{\sqrt{U^2-V^2}}\label{barU1}\,\,\,\\
&  \bar V =  f_H\left(\mathrm{sig}(U)\sqrt{h_{ER}(U^2-V^2)}\right)\frac{V}{\sqrt{U^2-V^2}} \ .\label{barV1}
\end{eqnarray}
Note that they coincide with the expressions of the map $P_1$, given by \eqref{co1}, if we define $S= f_{H}/D$. If we had chosen a different surface of revolution diffeomorphic to the hyperboloid, we would had obtained a different function $S(\cdot)$ with similar characteristics. That is why in the definition of $P_1$ we have generalized the function $S(\cdot)$ to any positive, smooth and monotonous increasing  function verifying $S(0)=1$. Therefore, any point of the manifold $ERb/\hspace{-0.17 cm}\sim$ defined with the coordinates $(\bar U_1,\bar V_1)$ (all points except the bridge), can be related with the hyperboloid (or a similar surface of revolution) by the diffeomorphism $\bar U_1=\bar U\,,\,\bar V_1=\bar V$, covering all the surface except the throat $\bar J$. The points of the bridge (and its neighbourhood) in the manifold $ERb/\hspace{-0.17 cm}\sim$ can be expressed with the coordinates $(\bar U_2,\bar V_2)$, defined in \eqref{U2V22}. The bridge correspond to the set $B =\{(\bar U_2,\bar V_2)\mid  \bar U_2^2 - \bar V_2^2= D^2\, , \bar U_2 > 0\}-(D,0)$, so it can be related to the throat of the hyperboloid by the simple diffeomorphism $\bar U_2=\bar U\,,\,\bar V_2=\bar V$. In the Proposition 2 we have proved that the coordinate systems $(\bar U_1,\bar V_1)$ and $(\bar U_2,\bar V_2)$ are compatible in the intersection of their domains, so there are no discontinuities. Therefore the manifold $ERb/\hspace{-0.17 cm}\sim$ is diffeomorphic to the covering space of the one sheet hyperboloid $\bar H$, after removing the point $(D,0)$.\\

\section*{Appendix B: Definitions} \label{definitions}

We shall summarize the main definitions used in the paper in order to facilitate the reading.\\

$ ERb = \{(U,V) \in \mathbb{R}^2 \mid U^2-V^2 \geq 0\} - (0,0)$ \\
 
$ \partial  ERb =  \{(U,V) \in \mathbb{R}^2 \mid U^2-V^2 =0\} - (0,0)$ \\

 $R_1\subset ERb =  \{(U,V) \in \mathbb{R}^2 \mid 0 < U^2-V^2 < \epsilon^2, U < 0 \} \}$ \\
 
 $R_2 \subset ERb =  \{(U,V) \in \mathbb{R}^2 \mid 0 < U^2-V^2 < \epsilon^2, U > 0 \} \}$ \\
 
 $ \Gamma_1 \subset ERb \, , \, \Gamma_1 =  \{(U,V) \in \mathbb{R}^2 \mid 0 < U^2-V^2 \} $\\
 
  $ \Gamma_2 \subset ERb \, , \, \Gamma_2 = \{(U,V) \in \mathbb{R}^2 \mid 0 \le U^2-V^2 < \epsilon^2 \} - (0,0)$ \\

 $\bar H = \{(\bar U , \bar V )\in \mathbb{R}^2  \mid  \bar U^2 - \bar V^2 > 0 \, , \bar U > 0\} $ \\

$ \bar J \subset \bar H = \{(\bar U , \bar V )\in \mathbb{R}^2  \mid  \bar U^2 - \bar V^2 = D^2 \, , \bar U > 0\} $ \\
 
 $ \bar H_1 \subset \bar H \, , \, \bar H_1 = \{(\bar U_1, \bar V_1 )\in \mathbb{R}^2  \mid   \bar U_1^2 - \bar V_1^2 > 0 \, , \bar U_1 > 0\} \setminus\bar J$ \\
 
    $ \bar H_2 \subset \bar H \, , \ \bar H_2 = \{(\bar U_2, \bar V_2 )\in \mathbb{R}^2  \mid \bar U_L(\bar V_2) <  \bar U_2 < \bar U_R(\bar V_2)\}- (D,0) $\\
    
 $ P_1 : \Gamma_1\subset ERb \rightarrow \bar H_1 \subset \bar H $\\
 
  $ P_2 : \Gamma_2\subset ERb \rightarrow  \bar H_2 \subset \bar H $\\

  $\bar R_1 =\{(\bar U_2, \bar V_2 )\in (\bar H_2 - \bar J ) \mid  \bar U_2 < \sqrt{D^2 + \bar V_2^2}\}$\\
  
  $\bar R_2 =\{(\bar U_2, \bar V_2 )\in (\bar H_2 - \bar J ) \mid  \bar U_2 > \sqrt{D^2 + \bar V_2^2}\}$\\

    $P_2(R_i) = \bar R_i \, , \ i = 1, 2$\\

$\varphi = \sqrt{\bar V_2^2 + D^2} - \bar U_2 $\\

\end{document}